\documentclass[3p,times]{elsarticle}

\usepackage{ecrc}

\usepackage{epstopdf}

\volume{00}

\firstpage{1}

\journalname{Nuclear Physics A}

\runauth{Xiaofeng Luo, Bedangadas Mohanty, Nu Xu}

\jid{nupha}

\jnltitlelogo{Nuclear Physics A}

\usepackage{graphicx}
\usepackage{amsmath,amssymb}

\newcommand{\sNN}{{{$\sqrt{s_{_{{NN}}}}$}}}

\newcommand{\KV}{{\mbox{$\kappa\sigma^{2}$}}}

\newcommand{\var}{{\mbox{$\sigma^{2}$}}}

\begin{document}

\begin{frontmatter}



\title{Baseline for the cumulants of net-proton distributions at STAR}



\author[label1,label4]{Xiaofeng Luo}
\author[label2]{Bedangadas Mohanty}
\author[label1,label3]{Nu Xu}
\address[label1]{Institute of Particle Physics and Key Laboratory of Quark \& Lepton Physics (MOE), \\Central China Normal University, Wuhan, 430079, China. }
\address[label2]{School of Physical Sciences, National Institute of Science Education and Research, Bhubaneswar 751005, India}
\address[label3]{Nuclear Science Division, Lawrence Berkeley National Laboratory, Berkeley, CA 94720, USA.}
\fntext[label4]{xfluo@mail.ccnu.edu.cn}

\begin{abstract}
We present a systematic comparison between the recently measured cumulants of the net-proton distributions by STAR for 0-5\% central Au+Au collisions at {\sNN}=7.7-200 GeV and two kinds of possible baseline measure, the Poisson and Binomial baselines. These baseline measures are assuming that the proton and anti-proton distributions independently follow Poisson statistics or Binomial statistics. The higher order cumulant net-proton data are observed to deviate from all the baseline measures studied at 19.6 and 27 GeV.  We also compare the net-proton with net-baryon fluctuations in UrQMD and AMPT model, and convert the net-proton fluctuations to net-baryon fluctuations in AMPT model by using a set of formula.
\end{abstract}
 

\end{frontmatter}



\section{Introduction}
\label{intro}
In the phase diagram of Quantum ChromoDynamics (QCD-theory of strong interactions), it is conjectured on the basis of theoretical calculations that there will be a QCD Critical Point (CP) at high temperature ($T$) and non-zero baryonic chemical potential ($\mu_{B}$)~\cite{QCP_Prediction}. Since the ab initio Lattice QCD calculation meet the notorious sign problem, there are still large uncertainties in theoretically determining  the location of the CP in the QCD phase diagram~\cite{qcp,qcp_Rajiv}.  Different QCD based models also give very different results~\cite{location}. Finding the existence of a CP experimentally will be an excellent test of QCD theory in the non-perturbative region and  a milestone of exploring the QCD phase diagram~\cite{science}. This is one of the main goals of the Beam Energy Scan (BES) Program at the Relativistic Heavy Ion Collider (RHIC). By tuning the colliding energy of gold nucleus from $\sqrt{s_{NN}}$=200 GeV down to 7.7 GeV, one can access a broad region of the QCD Phase Diagram (20$<\mu_{B}<$420 MeV)~\cite{bes}.  Due to the high sensitivity to the correlation length ($\xi$) of the dynamical system~\cite{qcp_signal,ratioCumulant,Neg_Kurtosis} and direct connection to the susceptibilities in theoretical calculations, for example, the Lattice QCD calculations~\cite{science,Lattice}, higher moments of multiplicity distributions of conserved quantities, such as net-baryon, net-charge and net-strangeness,  have been applied to search for the QCD critical point in heavy-ion collision experiments~\cite{highmoment}. As the correlation length will diverge near the CP, the non-monotonic variation of the moments of multiplicity distribution with respected to the colliding energy is the golden signature of the CP.  To extract the CP induced experimental signal, it is crucial to understand the non-CP physics effects in heavy-ion collisions on the experimental observable, such as the effects of conservations for charges (electric, baryon number and strangeness number), finite size, resonance decay and hadronic scattering. Moreover, proper baseline needs to be constructed for experimental observables to search for the CP.

In this paper, we will make the comparison between the baselines (Poisson, Binomial) and recently measured cumulants of net-proton distribution published by the STAR Collaboration and discuss the deviations of the data from the baselines. In addition, the results from the AMPT~\cite{ampt} and UrQMD~\cite{urqmd} models will be discussed.
\section{Results and Discussion}
Recently, the STAR Collaboration has published the beam energy dependence of higher moments of the net-proton (as proxy of net-baryon~\cite{Hatta}) multiplicity distributions from RHIC BES Au+Au collision data~\cite{STAR_BES_PRL}. The protons and anti-protons are identified with ionization energy loss in the Time Projection Chamber (TPC) of the STAR detector within transverse momentum range $0.4<p_{T}<0.8$ GeV/c and at mid-rapidity $|y|<0.5$. Several analysis techniques~\cite{technique,Delta_theory}  have been applied,  such as defining a new collision centrality with charged particle multiplicity within large pseudo-rapidity ($|\eta|<1$)  and excluding the particle of interest used in the analysis, centrality bin width correction (CBWC) and efficiency correction. These address the effects of auto-correlation, volume fluctuation and finite tracking efficiency in the moments calculation. 
\begin{figure}
\begin{center}
\includegraphics*[width=10.cm]{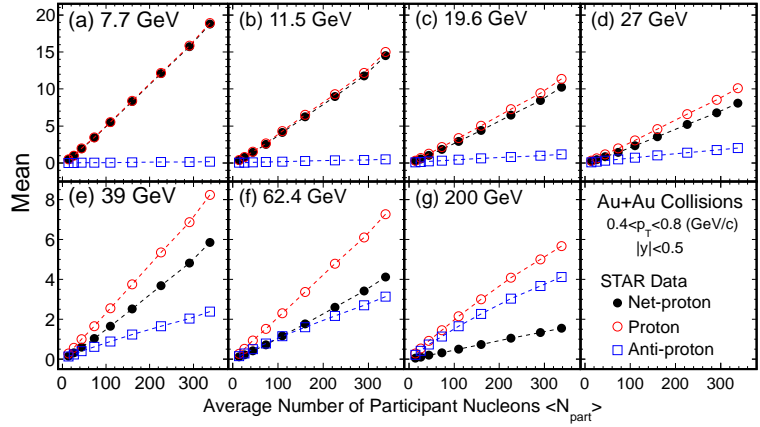}\\
\caption{ (Color Online) Efficiency corrected mean values of the net-proton, proton and anti-proton distributions  as a function of the average number of participant nucleons ($<N_{part}>$) in Au+Au collisions at $\sqrt{s_{NN}}$=7.7-200 GeV~\cite{STARwebpage}. The 
dashed lines are used to guide eyes.}
\label{fig:mean}
\end{center}
\end{figure}
Figure 1 shows the  mean of the net-proton, proton and anti-proton distributions are almost linearly increasing with the average number of participant nucleons for each energy. The different mean values of the proton and anti-proton distributions at each energy are determined by the interplay between baryon number pair production and baryon stopping effects. At high energies,  the pair production process dominates the production of protons and anti-protons at mid-rapidity, while at low energies the effect of baryon stopping is more important than at high energies.  In the following, we discuss some expectations  for cumulants of net-proton multiplicity distributions from some basic distributions. 
\begin{enumerate}
 \item {\bf Poisson :}
If the protons and anti-protons are independently distributed as Poissonian distributions. Then the net-proton multiplicity will follow the Skellam distribution, which is expressed as: \\ $P(N) = {(\frac{{{M_p}}}{{{M_{\overline p}}}})^{N/2}}{I_N}(2\sqrt {{M_p}{M_{\overline p}}} )\exp [ - ({M_p} + {M_{\overline p}})],$ where the $N$ is the net-proton number,  $I_{N}(x)$ is a modified Bessel function, $M_{p}$ and $M_{\overline p}$ are the mean number of protons and anti-protons, as shown in Fig. 1. The various order cumulants ($C_{n}$) are closely connected with the moments, e.g., $C_{1}=<N>=M,C_{2}=<(\Delta N)^2>=\sigma^{2}, C_{3}=<(\Delta N)^3>=S \sigma^{3}, C_{4}=<(\Delta N)^4>-3<(\Delta N)^2>^2=\kappa \sigma^{4}$, where the $\Delta N$ =$N-<N>$, the \var, $S$ and $\kappa$ are variance, skewness and kurtosis, respectively. Then we construct,  $S\sigma  = {C_3}/{C_2} = ({M_p} - {M_{\overline p }})/({M_p} + {M_{\overline p }})$ and $\kappa {\sigma ^2} = {C_4}/{C_2} = 1$, which provides the Poisson expectations for the 
 various order cumulants/moments of net-proton distributions. The only input parameters of the Poisson baseline for cumulants of net-proton distributions are the mean values of the proton and anti-proton distributions.

 \item {\bf Binomial/Negative Binomial:}
If the protons and anti-protons are independently distributed as Binomial or Negative Binomial distributions (BD/NBD). Then various order cumulants of the net-proton distributions can be expressed in term of cumulants of the proton and anti-proton distributions: $C_n^{net - p} = C_n^p + {( - 1)^n}C_n^{\bar p}$.
The first four order cumulants can be written as: 
$C_2^x = \sigma _x^2 = {\varepsilon _x}{\mu _x},C_3^x = {S_x}\sigma _x^3 = {\varepsilon _x}{\mu _x}(2{\varepsilon _x} - 1),C_4^x = {\kappa _x}\sigma _x^4 = {\varepsilon _x}{\mu _x}(6\varepsilon _x^2 - 6{\varepsilon _x} + 1)$
, where $\varepsilon_x=\sigma _x^2/\mu_x$, $\mu_x=M_x$, $M_{x}$ is the mean values of protons or anti-protons distributions, $x$=$p$ or $\bar{p}$. $\varepsilon_x<1$ means the underlying distributions of protons or anti-protons are Binomial distributions, 
while $\varepsilon_x>1$ gives Negative Binomial distributions. The input parameters for BD/NBD expectations are the measured mean and variance of the proton and anti-proton distributions.


 \end{enumerate}

\begin{figure}[htb]

\begin{minipage}[t]{0.5\linewidth}
\centering 
    \includegraphics[scale=0.33]{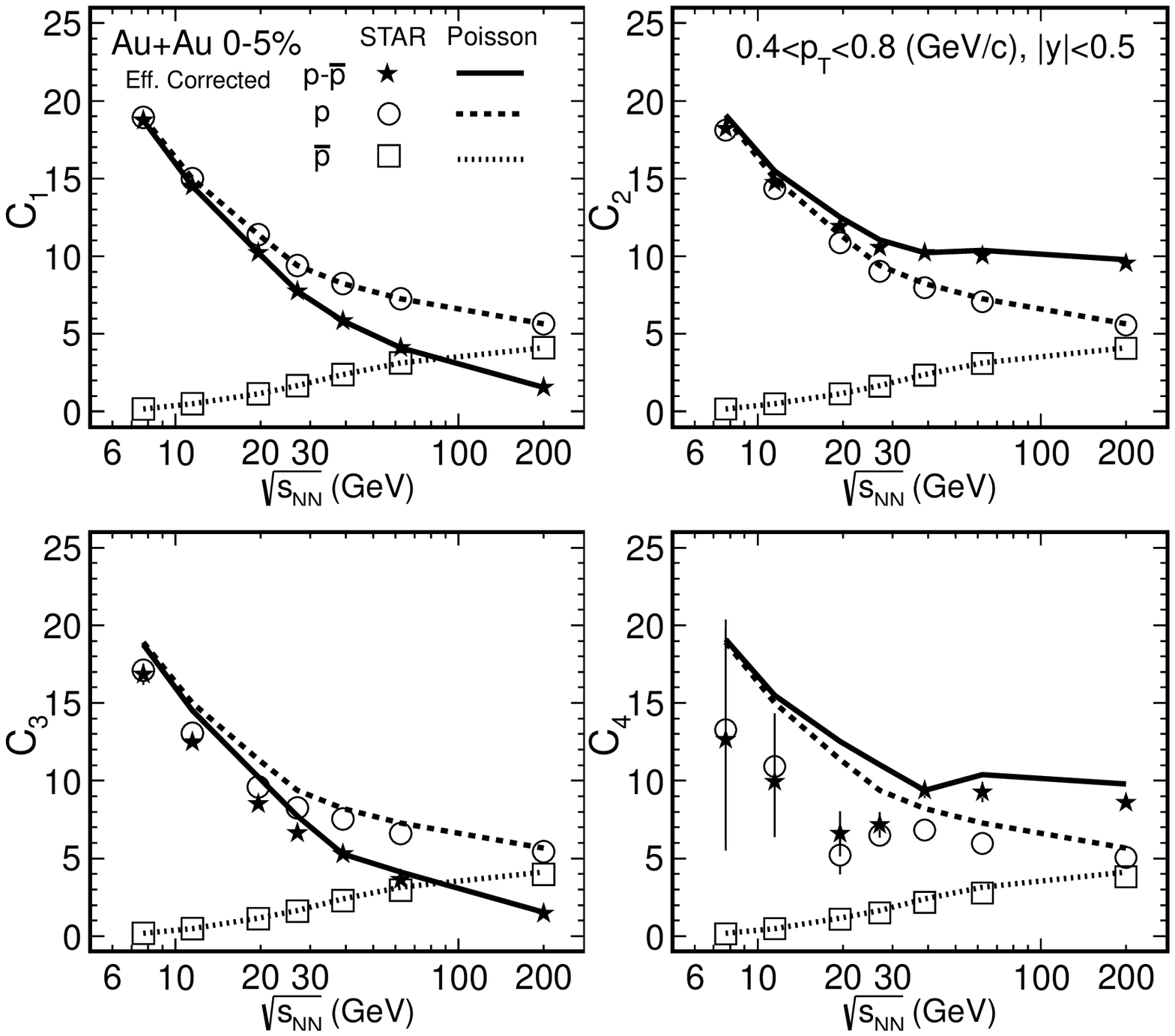}
    \end{minipage}
  \begin{minipage}[t]{0.5\linewidth}
  \centering 
   \includegraphics[scale=0.345]{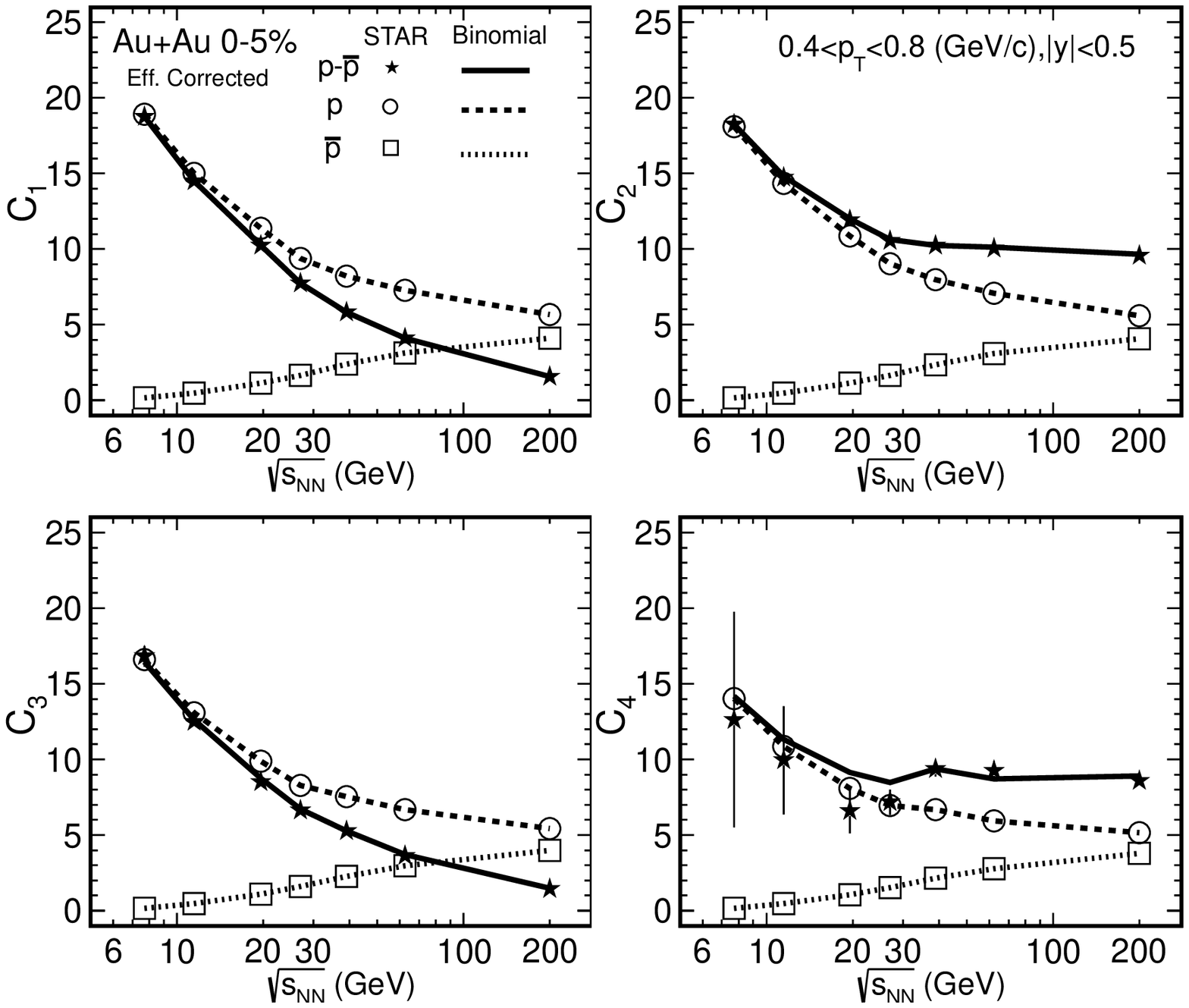}
      \end{minipage} 
      
      \caption{ Energy dependence of cumulants ($C_{1}-C_{4}$) of net-proton distributions for $0-5\%$ Au+Au collisions~\cite{STARwebpage}.
The error bars are statistical only. The dashed lines in the left panels are the Poisson expectations, while those are the Binomial expectations in the right panels. } \label{fig:cumulants_energy}

\end{figure}

Figure 2 shows the comparison between the cumulants of net-proton distributions and the Poisson/Binomial baselines. For the Poisson case, when the order of cumulant increases, the deviations of the data from the Poisson expectations for net-proton and proton increase. Largest deviations are found for $C_{4}$ at 19.6 and 27 GeV. For the Binomial case, the agreements persist up to the third order, while it fails to describe the net-proton and proton fourth order cumulants at 19.6 and 27 GeV.  It is understandable that the Binomial baseline describes the data better, as it uses the variance as an input parameter, in addition to the mean values, and as the variance is correlated with higher order cumulants. The cumulants of anti-proton distributions can be described by the Poisson and Binomial baselines very well. More baseline discussions from Hadronic Resonance Gas model and transport model UrQMD can be found in ~\cite{baseline_PRC}. The findings are consistent with ours. As we discussed that the binomial distributions can be used as baseline comparison for net-proton multiplicity distributions, an alternative interpretation, based on the negative binomial distribution, was given in ~\cite{westfall}. The negative binomial distributions used there is not appropriate for the net-proton multiplicity distributions.

The STAR experiment measures net-proton fluctuations instead of net-baryon fluctuations and one may want to know to what extend they can reflect the net-baryon fluctuations in heavy-ion collisions. Therefore, fig. 3 demonstrates the comparison between moments of net-proton and net-baryon distributions from AMPT and UrQMD model calculations. We can find that the {\KV} of net-baryon distributions are systematically lower than the net-proton results.  The differences are even bigger for low energies than high energies. There are two possible effects for the difference between net-proton and net-baryon fluctuations, one is the non inclusion of neutrons in the net-proton fluctuations, and the other one is the nucleon isospin exchanging process due to $\Delta$ resonance formation via $p\pi$ and $n\pi$ interaction, the so called isospin randomization, which will modify the net-proton fluctuations after the chemical freezeout. Asakawa and Kitazawa have derived a set of formulas~\cite{Asakawa_formula} to convert the measured net-proton cumulants to the net-baryon cumulants by taking into account the above two effects. The converting formulas for various order net-baryon cumulants can be written as:

\[\begin{array}{l}
\begin{array}{*{20}{c}}
{C_1^{net - B} = 2C_1^{net - p},}&{C_2^{net - B} = 4C_2^{net - p} - 2C_1^{tot - p},}
\end{array}\\
\begin{array}{*{20}{c}}
{C_3^{net - B} = 8C_3^{net - p} - 12(C_2^p - C_2^{\bar p}) + 6C_1^{net - p},}&{}
\end{array}\\
\begin{array}{*{20}{c}}
{C_4^{net - B} = 16C_4^{net - p} + 16C_3^{tot - p} - 64(C_3^p + C_3^{\bar p}) + 48C_2^{net - p} + 12C_2^{tot - p} - 26C_1^{tot - p},}&{}
\end{array}
\end{array}\]
where $tot$-$p$ means proton number plus anti-proton number. The right side of fig. 3 shows the net-baryon {\KV} ($C_{4}/C_{2}$) results, converted from the net-protons fluctuations. Within large uncertainties, they are consistent with the net-baryon results directly calculated with the AMPT model.
\begin{figure}[htb]

\begin{minipage}[c]{0.5\linewidth}
\centering 
    \includegraphics[scale=0.47]{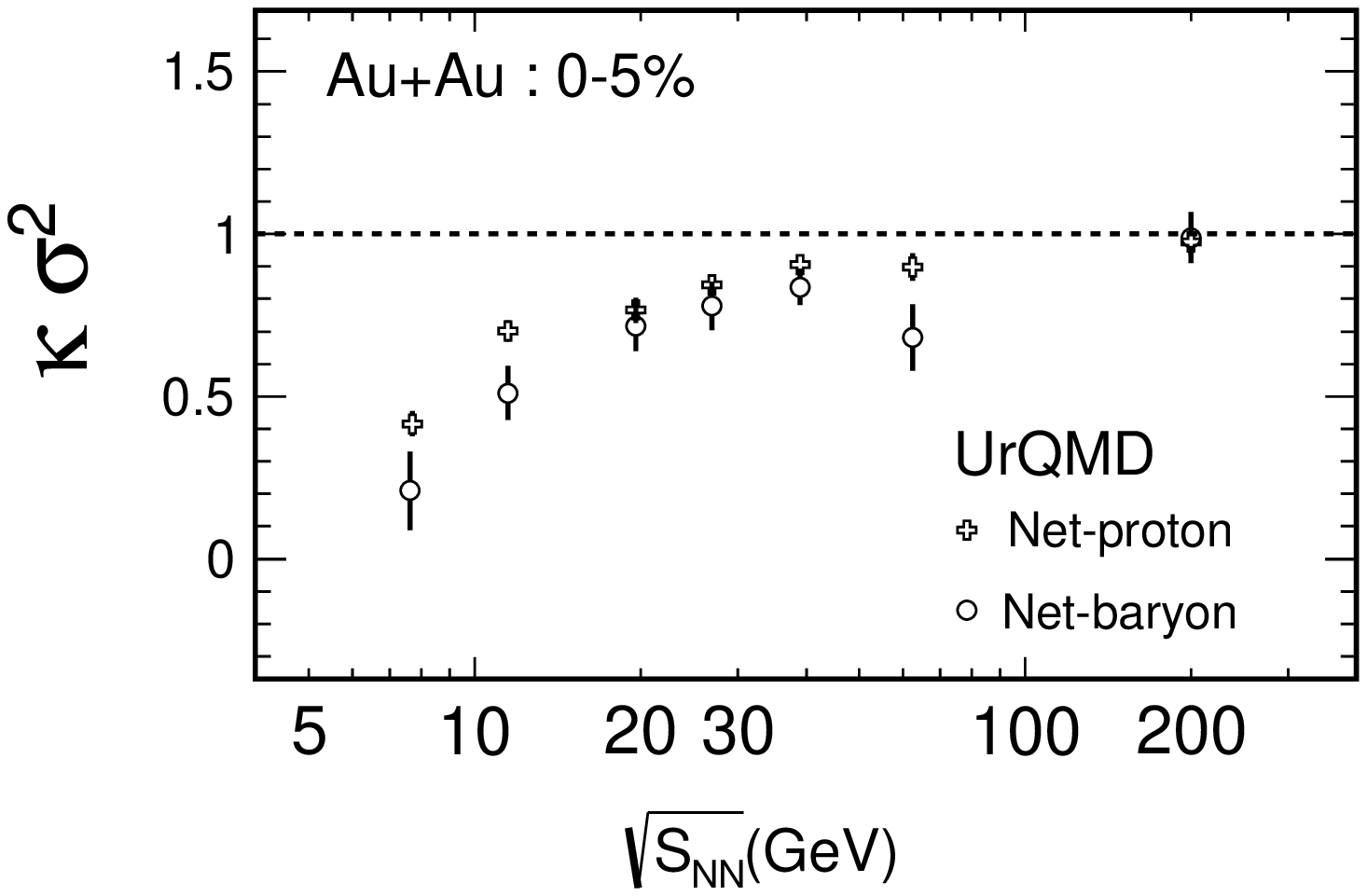}
    \end{minipage}
  \begin{minipage}[c]{0.5\linewidth}
  \centering 
   \includegraphics[scale=0.45]{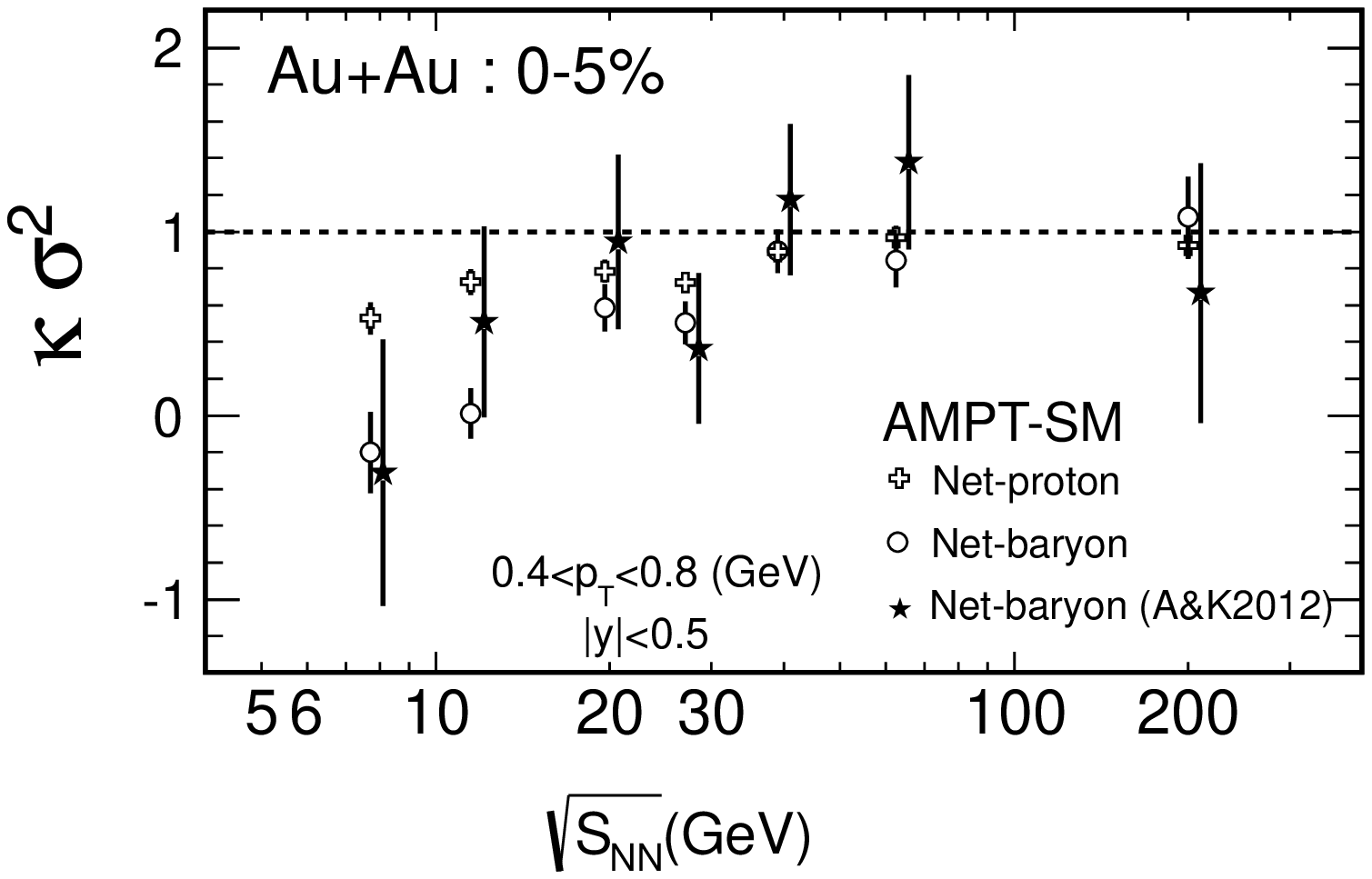}
      \end{minipage} 
      
      \caption{Energy dependence of {\KV} of net-proton and net-baryon distributions for 0-5\% Au+Au collisions from the UrQMD (left) and
    the AMPT string melting model (right). The results marked as solid black stars are based on theoretical calculations using Asakawa 
    and Kitazawa's formula. The error calculation is based on the Bootstrap method. } \label{fig:MP_energy}
\end{figure}

\section{Summary}
We have compared the energy dependence of cumulants of net-proton distributions in 0-5\% central Au+Au collisions with Poisson and Binomial baselines. The Binomial baseline describes the data better than the Poisson baseline. The largest deviation is observed for the fourth order net-proton cumulant ($C_{4}$) from Poisson and Binomial baselines in the most central collisions at 19.6 and 27 GeV. A proper comparison of the experimental measurements to QCD calculations is needed to extract the exact physics process that leads to deviation of the data from the baselines presented. Differences between net-proton and net-baryon fluctuations are observed in the AMPT and UrQMD models. Finally, we use the A\&K's formulas to convert the net-proton fluctuations in the AMPT model to net-baryon fluctuations, which is consistent with the net-baryon results directly calculated from the AMPT model.

\section*{Acknowledgement}
The work was supported in part by the MoST of China 973-Project No.2015CB856901, NSFC under grant No. 11205067, 11221504 and 11228513.  China Postdoctoral Science Foundation (2012M511237, 2013T60732).








\bibliography{QM14_XiaofengLuo}
\bibliographystyle{unsrt}

\end{document}